\documentclass[useAMS,usegraphicx]{mn2e}

\usepackage{multirow} 
\usepackage{epsfig}
\usepackage{subfigure}
\usepackage[all]{xy}
\usepackage{float}
\usepackage{lscape}  

\newcommand{\etal}{et al.}

\def\mnras{MNRAS}
\def\apj{ApJ}
\def\apjs{ApJS}
\def\aap{A\&A}

\def\apjl{ApJ Lett.}

\title[Multi-epoch analysis of ESO 362--G18] {Black hole spin and size of the X-ray emitting
  region(s) in the Seyfert~1.5 galaxy ESO 362--G18}
\author[B.~Ag\'is-Gonz\'alez \etal]
       {\parbox{\textwidth}{B.~Ag\'is-Gonz\'alez,$^1$\thanks{E-mail:
             \texttt{agisgb@cab.inta-csic.es}} G.~Miniutti,$^1$
           E.~Kara,$^2$ A.C.~Fabian,$^2$ M.~Sanfrutos,$^1$
           G.~Risaliti,$^{3,4}$ S.~Bianchi,$^5$ 
           N.L.~Strotjohann,$^6$ R.~D.~Saxton$^7$ and M.L.~Parker$^2$
         } \vspace{0.5cm}\\
\parbox{\textwidth}{ 
$^1$ Centro de Astrobiologia (CSIC--INTA),
  Dep. de Astrofisica; ESA, P.O: Box 78, E-28691, Villanueva de la
  Ca\~nada, Madrid, Spain \\ $^2$ Institute of Astronomy, Madingley
  Road, Cambridge CB3~0HA \\ $^3$ Harvard--Smithsonian Center for
  Astrophysics, 60 Garden St., Cambridge MA 02138, USA \\ $^4$ INAF -
  Osservatorio Astronomico di Arcetri, L.go E. Fermi 5, Firenze, Italy
  \\ $^5$ Dipartimento di Matematica e Fisica,
  Universit\`a degli Studi di Roma Tre, via della Vasca Navale 84, 00146 Roma, Italy
\\ $^6$ Physikalisches Institut, Universit\"at Bonn, Nussallee 12, D-53115 Bonn, Germany
  \\ $^7$ XMM SOC, ESA, P.O. Boc 78, E-28691, Villanueva de la
  Ca\~nada, Madrid, Spain  }}

\pagerange{\pageref{firstpage}--\pageref{lastpage}}
\pubyear{2001}

\usepackage{times}

\begin{document}

\label{firstpage}

 \maketitle

\begin{abstract}

We report results from multi--epoch X--ray observations of the
Seyfert~1.5 galaxy ESO~362--G18 performed between November 2005 and
June 2010. ESO~362--G18 generally exhibits the typical X--ray spectrum
of type 1 Active Galactic Nuclei (AGN). A disc--reflection component
accounts for broad residuals in the iron K band and above 10~keV, as
well as for a significant soft excess.  From our best--fitting
reflection model, we measure a black hole spin $a\geq 0.92$ at the
99.99 per cent confidence level. ESO~362--G18 is also (typically)
mildly absorbed by a column of neutral gas. The absorber is variable
and one observation, performed $\sim 2$~months after a typical mildly
absorbed one, is heavily absorbed by a cold column density of $\sim
3-4\times 10^{23}$~cm$^{-2}$, nearly two orders of magnitude higher
than that during any other observation. UV variability between the
heavily absorbed observation and the others suggests that the absorber
can be identified with a dusty, clumpy torus. The absorption
variability timescale enables us to locate the X--ray emitting region
within the innermost $\sim 50$ gravitational radii. Such result holds
not only for the X--ray continuum, but also for the soft excess.
\end{abstract}


\section{Introduction}

\begin{table*}
\caption{Details of the X--ray observations used in this work. The
  last two columns refer to the X--ray counts collected in a soft and
  hard energy band. The soft band is the 0.3--2~keV for the {\it
    Swift}, XMM~1 and XMM~2 observation, and the 0.5--2~keV for the
  {\it Suzaku} and the five {\it Chandra} observations. The hard band
  lower limit is set at 2~keV, while the upper bound is 10~keV except
  for the {\it Swift} and the five {\it Chandra} observations, where
  we use data up to 8.5~keV and 7.5~keV respectively because of
  signal--to--noise limitations.}
\label{obslog} 
\begin{center}
\begin{tabular}{l c c c c c c c }
\hline\hline 
Obs.  & Obs. ID & Obs. date  &   Net exp. & $F_{0.5-2}$ & $F_{2-10}$ & Soft counts& Hard counts\\
\\
Swift        &  00035234002  &  2005-11-26 & $7$  & $3.9\pm 0.1$     & $13.5\pm 0.4$   & $1.7$ & $1.0$ \\
XMM~1        &  0312190701   &  2006-01-28 & $8$  & $0.66\pm 0.01$   & $3.7\pm 0.1$    & $4.3$ & $2.1$ \\
Suzaku       &  703013010    &  2008-04-11 & $41$ & $27.2\pm 0.1$    & $33.2\pm 0.1$   & $127.4$ & $76.4$ \\
XMM~2        &  0610180101   &  2010-01-29 & $53$ & $2.86\pm 0.02$   & $9.58\pm 0.04$  & $107.8$ & $43.6$ \\
Chandra~1    &  11608        &  2010-05-18 & $10$ & $2.67\pm 0.05$ & $10.5\pm 0.2$  & $6.5$ & $3.9$ \\
Chandra~2    &  11609        &  2010-05-19 & $10$ & $4.29\pm 0.06$ & $13.3\pm 0.2$  & $10.4$ & $5.1$ \\
Chandra~3    &  11610        &  2010-05-21 & $10$ & $2.11\pm 0.04$ & $6.5\pm 0.2$  & $5.3$ & $2.5$ \\
Chandra~4    &  11611        &  2010-05-25 & $10$ & $10.2\pm 0.1$  & $19.9\pm 0.2$  & $24.0$ & $8.6$ \\
Chandra~5    &  11612        &  2010-06-03 & $10$ & $6.55\pm 0.07$ & $14.9\pm 0.2$  & $15.2$ & $5.8$ \\
\hline     
\hline
\end{tabular}
\end{center}
\raggedright The net exposure (fourth column) is in units of
ks. Observed fluxes in the 0.5--2~keV (fifth column) and 2--10~keV
bands (sixth column) are given in units of
$10^{-12}$~erg~s$^{-1}$~cm$^{-2}$. The soft and hard X--ray counts
(last two columns) are in units of $10^3$~photons.
\end{table*}

In recent years, several examples of X--ray absorption variability in
Active Galactic Nuclei (AGN) have been reported on relatively short
timescales ranging from hours to days (e.g. NGC~4388, Elvis et
al. 2004; NGC~4151, Puccetti et al. 2007; NGC~1365, Risaliti et
al. 2005; 2007; 2009 and Maiolino et al. 2010; NGC~7582, Bianchi et
al. 2009; Mrk~766, Risaliti et al. 2011; SWIFT~J2127.4+5654, Sanfrutos
et al. 2013). The data are generally consistent with absorption by
neutral/low--ionization clouds with typical column density of
$10^{22}-10^{24}$~cm$^{-2}$, number density of
$10^9-10^{11}$~cm$^{-3}$, and velocity of few times
$10^{3}$~km~s$^{-1}$ which suggests that the variable X--ray absorber
can be identified with broad line region (BLR) clouds. In the
best studied cases, occultation events have been followed almost
entirely from ingress to egress enabling e.g. Risaliti et al. (2007,
2009 in NGC~1365) and Sanfrutos et al. (2013 in SWIFT~J2127.4+5654) to
infer that the X--ray source is compact, and confined within
$5-10~r_g$ ($1~r_g = G~M_{\rm{BH}}/c^2$) around the central black
hole.

X-ray absorption variability is also seen on longer timescales. These
events are obviously more difficult to detect, as the longer timescale
ideally requires long and continuous X--ray monitoring campaigns. In
the few cases reported so far (e.g. Cen~A, Rivers, Markowitz \&
Rothschild 2011: NGC~4507, Marinucci et al. 2012; ESO~323--G77,
Miniutti et al. 2014; see Markowitz, Krumpe \& NIkutta 2014 for a
recent statistical analysis on such events) absorption on timescales
of a few months can be attributed to clouds with typical column
density of $10^{22}-10^{23}$~cm$^{-2}$ and number density of
$10^7-10^8$~cm$^{-2}$, which points towards a clumpy torus origin
(Nenkova et al. 2008a, 2008b).

Here we discuss the case of ESO~362--G18 (a.k.a. MCG~05--13--17), a
Seyfert~1.5 galaxy at redshift $z=0.012$ (e.g. Bennert et
al. 2006). The X--ray luminosity from a {\it ROSAT} observation in the
early 90s is $\sim 4\times 10^{41}$~erg~s$^{-1}$ in the 0.1-2.4~keV
band, much lower than that required to model the extended narrow line
region (ENLR) properties (Fraquelli, Storchi-Bergmann \& Binette
2000). This suggests that the {\it ROSAT} observation may have been
affected by X--ray absorption, with different soft X--ray fluxes into
our line--of--sight and into that of the ENLR. Subsequent observations
with {\it Swift} and {\it XMM--Newton} confirm that X--ray absorption
variability is present towards the X--ray--emitting region of
ESO~362--G18, and that the soft X--ray luminosity during unabsorbed
states is much higher than that derived by {\it ROSAT}. In fact,
ESO~362--G18 makes a transition from an unobscured state with soft X--ray
luminosity of $L_{0.5-2} \sim {\rm{few}}\times 10^{42}$~erg~s$^{-1}$
observed with {\it Swift} to a highly absorbed state with one order of
magnitude lower soft luminosity observed with {\it XMM--Newton} about
2 months later (Winter et al. 2008).

We consider here 9 X--ray observations of ESO~362--G18 performed with the
{\it Swift}, {\it XMM--Newton}, {\it Suzaku}, and {\it Chandra} X--ray
observatories. The first 2 observations were performed only $\sim
2$~months apart (2005-11-26 and 2006-01-28 respectively), and the
third was performed about $2.2$~years later. The remaining 6
observations were all performed within one year (2010) in the
framework of our dedicated X--ray monitoring of the source. In
particular, the 5 {\it Chandra} monitoring observations were performed
within 15 days, between 2010--05--18 and 2010--06--03. The most
important details of the observations used in this work are given in
Table~\ref{obslog}. Results from further X--ray observations performed
with {\it Swift} starting from November 2010 are reported elsewhere
(Miniutti et al. in preparation), as they provide interesting,
independent results by themselves.

\section{X-ray Observations}
\label{obs}

The data from the various missions and detectors have been reduced as
standard using the dedicated software {\small SAS v12.01} ({\it
  XMM--Newton}), {\small CIAO v4.4} ({\it Chandra}), and {\small
  HEASOFT v6.11} ({\it Swift} and {\it Suzaku}). Epoch-- and
position--dependent ancillary responses and redistribution matrices
have been generated for each data set. Source products have been
extracted from circular regions centered on the source, and background
products have been generated from nearby source--free regions. For
simplicity, we only consider here EPIC pn spectra from {\it
  XMM--Newton}, although we have checked their consistency with the
MOS data. As for the {\it Suzaku} data, we merge the spectra from the
front--illuminated CCD detectors XIS0 and XIS3 using the {\small FTOOL
  ADDASCASPEC} to obtain one single front--illuminated CCD spectrum of
the source. The resulting spectrum from the front--illuminated
detectors is consistent with that from the back--illuminated XIS1
detector, although only the former is used here for simplicity. We
also make use of the PIN data from the Suzaku HXD, which have been
reduced with the dedicated {FTOOL HXDGSOXBPI}. The source flux
obtained from the PIN detector is about 25 per cent of the background
below 40~keV. Our {\it Chandra} observations were performed in
continuous--clock mode to minimize pile--up and the resulting spectra
are indeed pile--up free. When needed to convert fluxes into
luminosities, we adopt a cosmology with
H$_0=70$~km~s$^{-1}$~Mpc$^{-1}$, $\Omega_\Lambda= 0.73$, and $\Omega_M
= 0.27$. Quoted uncertainties refer to the 90 per cent confidence
level for one interesting parameter unless otherwise stated.

\section{X--ray spectral variability}

As the five {\it Chandra} observations are part of a two--week long
monitoring campaign, we shall treat them separately from the others to
investigate any short--timescale spectral variability in detail (see
Section~\ref{chandra}). In order to first define a global X--ray
spectral model to be used to study the spectral variability of
ESO~362--G18, we start our analysis from the two highest--quality
observations of our campaign, namely the {\it Suzaku} and the XMM~2
observations which provide, by far, the largest number of X--ray
counts (see Table~\ref{obslog}).

\subsection{The high--resolution RGS data from the XMM~2 observation}

\begin{figure}
\begin{center}
\includegraphics[width=0.33\textwidth,height=0.45\textwidth,angle=-90]{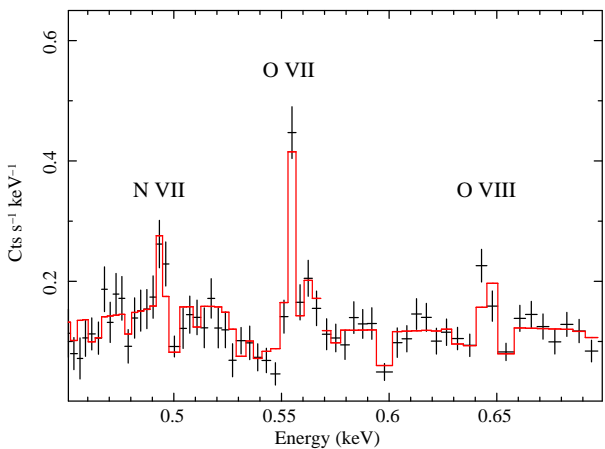}
\caption{The RGS~1 spectrum obtained during the XMM~2 observation is
  shown together with the best--fitting model. The most important
  emission lines are labeled for reference (the O~\textsc{vii} line
  being in fact a triplet). Data have been rebinned for visual
  clarity.}
\label{rgsspec}
\end{center}
\end{figure}

As a first step, we examine the high--resolution spectrum from the RGS
on board of {\it XMM--Newton} during the XMM~2
observation\footnote{The RGS data during the XMM~1 observation have
  poor quality due to the short exposure ($\sim 8$~ks) and relatively
  low soft X--ray flux ($F_{0.5-2}\sim 6.4\times
  10^{-13}$~erg~s$^{-1}$~cm$^{-2}$). On the other hand, the RGS data
  during the longer and higher--flux XMM~2 observation are of good
  enough quality to be used.}. We fit the RGS data with a simple
blackbody plus power law and Galactic absorption model ($N_{\rm
  H}=1.75\times 10^{20}$~cm$^{-2}$, Kalberla et al. 2005) which
is adequate for the continuum. Several soft X--ray narrow emission
lines are visually clear. We then add a series of Gaussian emission
lines, as required by the data. We detect 5 soft X--ray emission lines
corresponding to N~\textsc{vii} and O~\textsc{viii} Ly$\alpha$, and to
the O~\textsc{vii} triplet. The properties of the detected lines are
reported in Table~\ref{RGS}. In Fig.~\ref{rgsspec}, we show the most
relevant portion of the RGS~1 spectrum together with the best--fitting
model and the identification of the N~\textsc{vii}, O~\textsc{vii}
(actually a triplet), and O~\textsc{viii} emission lines. The
dominance of the forbidden O~\textsc{vii} Ly$\alpha$ line at 0.561~keV
strongly suggests emission by photo--ionized gas (e.g. Porquet \&
Dubau 2000). All subsequent spectral models include the emission lines
of Table~\ref{RGS} with energy fixed at the laboratory value, and with
normalization allowed to vary only within the uncertainties derived
from the RGS analysis.

\begin{table}
\caption{The soft X--ray emission lines detected in the RGS spectrum
  of the XMM~2 observation. Line energies are given in keV, and
  intensities are reported in units of $10^{-5}$~ph~s$^{-1}$~cm$^{-2}$.}
\label{RGS}      
\begin{center}
\begin{tabular}{l c c c}
\hline\hline                 
Line ID & Lab. Energy & Observables & Measurements \\
\hline
N~\textsc{vii} Ly$\alpha$ & 0.500    &Energy            &$ 0.49\pm 0.02$ \\  
                  &              &Intensity          &$1.3\pm 0.6$\\
\hline
                &  0.561 (f)   &Energy           &$ 0.561 \pm 0.001$ \\  
                   &               &Intensity           &$ 4.6\pm 1.0$\\

O~\textsc{vii} Ly$\alpha$  &  0.569  (i)   &Energy            &$ 0.569\pm 0.002$ \\  
                    &              &Intensity           &$1.2\pm 0.7$\\

                &  0.574 (r)   &Energy         &$ 0.575\pm 0.002$\\  
                &             &Intensity           &$0.7^{+1.0}_{-0.5}$\\
\hline
O~\textsc{viii} Ly$\alpha$ & 0.654    &Energy         &$ 0.653\pm 0.002$ \\
                      &         &Intensity        &$1.3\pm 0.5$\\ 
\hline
\hline                           
\end{tabular}
\end{center}
\end{table}

\subsection{The high-quality {\it Suzaku} and XMM~2 observations}

\begin{figure}
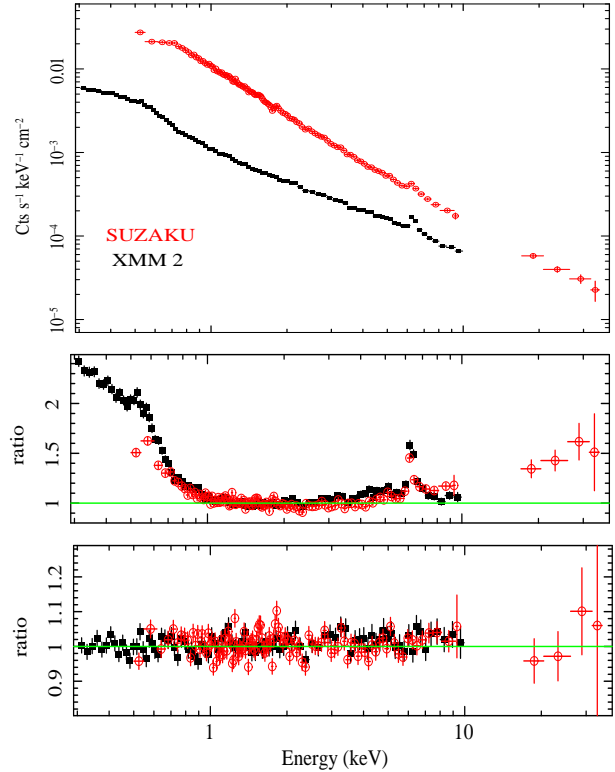

\begin{center}
\includegraphics[width=0.25\textwidth,height=0.45\textwidth,angle=-90]{HQdata.ps}
{\vspace{0.2cm}}
\includegraphics[width=0.13\textwidth,height=0.45\textwidth,angle=-90]{HQratioa.ps}
{\vspace{0.2cm}}
\includegraphics[width=0.17\textwidth,height=0.45\textwidth,angle=-90]{HQratio.ps}
\caption{In the upper panel, we show the high--quality data from {\it
    Suzaku} (upper open circles, red in the on--line version) and
  XMM~2 (lower, filled squares). The two spectra are normalized to the
  different detector effective area to show the real spectral shape.
  In the middle panel, we show the data--to--model ratios obtained by
  fitting a power law absorbed by the Galactic column density in the
  1--4~keV and 7--10~keV band (i.e. ignoring the soft excess, Fe K
  band, and hard X--ray data above 10~keV). This shows that both
  observations are characterized by a soft excess and exhibit broad
  residuals in the Fe K band. Moreover, the HXD/PIN data suggest the
  presence of a hard X--ray excess . In the lower panel, we show our
  final, broadband, best--fitting model, see text for details.}
\label{HQspec}
\end{center}
\end{figure}

The X--ray spectra from the two high--quality observations ({\it
  Suzaku} and XMM~2) are shown in the upper panel of
Fig.~\ref{HQspec}. In order to show the general spectral shape of the
two observations, we fit a power law plus Galactic absorption model in
the 1--4~keV and 7--10~keV bands, i.e. ignoring the soft and Fe K
bands, as well as the {\it Suzaku} HXD data above 10~keV. The
resulting data--to--model ratios are shown in the middle panel of
Fig.~\ref{HQspec}. A soft excess with similar spectral shape is
detected in both observations. Both spectra also exhibit broad
residuals in the Fe K band, and the HXD/PIN data
suggest that a hard X--ray excess is present above 10~keV. Note,
however, that while the best--fitting photon index from the (higher
flux) {\it Suzaku} observation is $\Gamma \simeq 2.0$, and in line with the typical spectral
slope of unobscured AGN ($< \Gamma> \simeq 1.9$, e.g. Piconcelli et
al. 2005), the XMM~2 observation is extremely hard with $\Gamma \simeq
1.3$, possibly signaling that some extra component is needed in the
hard band.

\begin{figure}
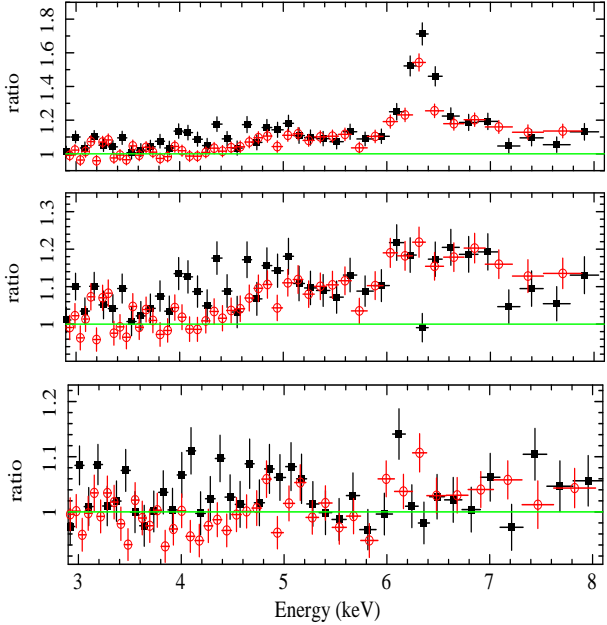

\begin{center}
{\includegraphics[width=0.13\textwidth,height=0.45\textwidth,angle=-90]{HQFea.ps}}
{\vspace{0.2cm}}
{\includegraphics[width=0.13\textwidth,height=0.45\textwidth,angle=-90]{HQFeb.ps}}
{\vspace{0.2cm}}
{\includegraphics[width=0.18\textwidth,height=0.45\textwidth,angle=-90]{HQFec.ps}}
\caption{In the upper panel, we show the data--to--model ratio for the
  {\it Suzaku} and the XMM~2 observations in the Fe K band for a
  simple power law model fitted in the 2--10~keV band only and
  ignoring the 4--7~keV energy range (symbols are the same as in
  Fig.~\ref{HQspec}). In the middle panel, we account for the presence
  of an unresolved Fe emission line with a simple Gaussian model. A
  broad feature is left in the Fe K band in both observations. The feature is accounted for by an additional relativistic emission line model (lower panel).}
\label{HQFe}
\end{center}
\end{figure}

As a first step, we examine in some more detail the Fe K band
spectra. Fitting a simple power law model (and Galactic absorption) in
the 2--10~keV band, and ignoring the 4--7~keV energy range where Fe
features are expected, produces the data--to--model ratio shown in the
upper panel of Fig.~\ref{HQFe}. Adding an unresolved Gaussian with
energy at 6.4~keV in the rest--frame accounts for the narrow Fe
emission line and leaves a broad feature in the Fe K band (middle
panel of Fig.~\ref{HQFe}). We then add a relativistic emission line
model (the {\small{LAOR}} model, Laor 1991), and the data are now well
reproduced in the 2--10~keV band with $\chi^2=1930$ for 1855 degrees of freedom (dof)
(lower panel of Fig.~\ref{HQFe}). The narrow Fe emission line has an
equivalent width of $\sim 40$~eV ($\sim 130$~eV), and the broad
relativistic one of $\sim 130$~eV ($\sim 250$~eV) in the {\it Suzaku}
(XMM~2) observation. As for the relativistic parameters (forced to be
the same in the two observations), we obtain a disc inclination of
$47^\circ\pm 8^\circ$, an emissivity index $q = 4.2 \pm 1.4$, and an
inner disc radius of $\leq 3.1~r_g$ that, when identified with the
innermost stable circular orbit around the black hole, corresponds to
a black hole spin of $a \geq 0.76$. However, describing the data in a
limited energy band with models that do not account for the reflection
continuum associated with the emission lines may lead to misleading
and/or inaccurate results. We then consider below a broadband spectral
analysis using more self--consistent, physically--motivated spectral
models.

Our broadband spectral model for both observations initially includes
(i) a power law continuum, (ii) a distant reflection component (we use
the {\small{PEXMON}} model in {\small{XSPEC}}, Nandra et al. 2007),
(iii) all soft X--ray emission lines of Table~\ref{RGS}, (iv) a
disc--reflection component which may account for the soft excess, and
for the positive residuals in the Fe K band (and in the HXD/PIN). The
latter is modeled with the Ross \& Fabian (2005) partially ionized
reflection model\footnote{We extend their original grid to include
  values of the ionization parameter down to $\xi =
  0.01$~erg~s$^{-1}$~cm and photon indices down to $\Gamma = 1.5$.}
which is convolved with the {\small KERRCONV} relativistic kernel
(Brenneman \& Reynolds 2006) to account for any relativistic
effects. The soft X--ray emission lines and the distant reflection
component are forced to be the same in both spectra, as these
components are not expected to vary.

The fit is relatively good with $\chi^2 = 3040$ for 2580 degrees of
freedom (dof), and the disc--reflection component accounts well for
all three features noted above, namely the soft excess, broad Fe K
residuals, and hard X--ray excess. The description of the data
improves by adding a rather typical warm absorber (we use the
{\small{ZXIPCF}} model in {\small{XSPEC}}) affecting both the power
law and the disc--reflection component. As the column density of the
warm absorber is consistent with being the same in the two
observations, we force a common $N_{\rm H}$ in both data sets. We
obtain a significant improvement to $\chi^2 = 2880$ for 2577 dof with
a common column density of $\simeq 10^{21}$~cm$^{-2}$ and $\log\xi
\simeq 2.4 $ ($\simeq 2.0$) for the {\it Suzaku} (XMM~2)
observation. Although the fit is now acceptable, the XMM~2 data are
not well reproduced with systematic positive residuals in the 2--6~keV
band. Moreover, the resulting spectral shape for that observation is
still rather flat ($\Gamma \simeq 1.5-1.6$) when compared with the
typical value for unobscured AGN ($<\Gamma> \sim 1.9$).

We then consider the possible presence of absorption by including a
neutral absorber affecting both the power law and disc--reflection
components. For the sake of generality, we allow it to partially cover
the X--ray source. The fit improves dramatically by
$\Delta\chi^2=-137$ for 4 more free parameters ($\chi^2=2743$ for 2573
dof). In fact, the {\it Suzaku} spectrum is consistent with being
unabsorbed (with covering fraction $\leq 0.1$). Hence, the improvement
is only due to a better description of the XMM~2 spectrum which
requires a neutral absorber with column density $N_{\rm H} \simeq
10^{22}$~cm$^{-2}$ and covering fraction $C_{\rm f} \simeq 0.4$. The
photon index during the XMM~2 observation is now $\Gamma \simeq 1.8$,
broadly consistent with a standard spectral slope. Replacing the
disc--reflection component with a phenomenological blackbody (or disc
blackbody) model to account for the soft excess results in a much
worse statistical description of the data ($\chi^2= 2938$ for 2576
dof), demonstrating that the disc--reflection model is a better
description for the soft excess.

\begin{figure}
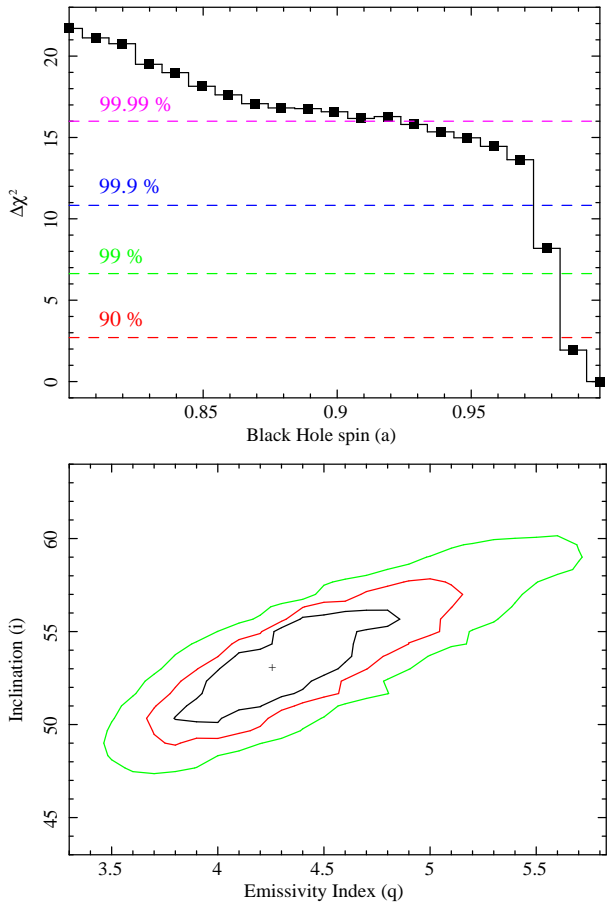

\begin{center}
\includegraphics[width=0.33\textwidth,height=0.45\textwidth,angle=-90]{spin.ps}
{\vspace{0.2cm}}
\includegraphics[width=0.33\textwidth,height=0.45\textwidth,angle=-90]{IQ.ps}
\caption{In the upper panel, we show the error contour for the black
  hole spin in ESO~362--G18, as obtained from the disc reflection
  model applied to the two high-quality observations ({\it Suzaku}
  and XMM~2). The contour has been obtained allowing
  all other model parameters to vary. The horizontal dashed lines are,
  from bottom to top, the 90, 99, 99.9, and 99.99 per cent confidence
  levels. Values of $a\leq 0.8$ always give $\Delta\chi^2 \geq 22$, and
  are not shown for visual clarity. In the lower panel, we show the 68,
  90, and 99 per cent confidence level contours for the
  disc--reflection emissivity index and for the disc inclination. The
  contours have been obtained allowing all other parameters to vary.}
\label{RelParams}
\end{center}
\end{figure}

Let us discuss here results on the relativistic parameters associated
with the disc--reflection component, namely the black hole spin $a$,
emissivity index $q$, and disc inclination $i$. The error contour for
the black hole spin is shown in the upper panel of
Fig.~\ref{RelParams}. We measure a black hole spin of $a\geq 0.92$ at
the 99.99 per cent confidence level, and of $\geq 0.98$ at the 90 per
cent one. Hence, the accreting black hole powering ESO~362--G18 is a
very rapidly spinning Kerr black hole, consistent with being maximally
spinning ($a=0.998$). The contours plot for the disc--reflection
emissivity index ($q$) and for the disc inclination ($i$) is shown in
the lower panel of the same Figure and indicates that
$q=4.3^{+0.8}_{-0.6}$ and $i=53^\circ\pm 5^\circ$ (90 per cent
confidence level for the two interesting parameters). All three
parameters are consistent with those derived from the simpler analysis
discussed above and performed with emission line models only (see
Fig.~\ref{HQFe} and the associated discussion). However, using the
broadband data and more self--consistent and physically--motivated
models significantly reduces the uncertainties on these relevant
parameters.

\subsection{Joint fits to the {\it Swift}, {\it Suzaku} and the two {\it XMM--Newton} spectra}
\label{HQ4sec}

Having found a good global spectral model for the two highest--quality
observations of our campaign, we now use it to perform a
simultaneous analysis of the first four X--ray observations of
Table~\ref{obslog}, namely to the {\it Swift}, XMM~1, {\it Suzaku},
and XMM~2 observations. We defer the analysis of the five {\it
  Chandra} monitoring observations to Section~\ref{chandra}. The
X--ray spectrum of ESO~362--G18 from the first four 
observations is shown in the upper panel of Fig.~\ref{HQ4} together
with the best--fitting models (see below). Remarkable spectral
variability is present with a general steeper--when--brighter
behavior. Moreover the (lowest flux) XMM~1 observation appears to
have a distinct spectral shape. The visually larger narrow Fe line
equivalent width suggests that the hard X--ray continuum is
significantly depressed during the XMM~1 observation, either because
of an intrinsic low flux state or because of absorption.

We proceed by performing a joint fit to the four observations with the
same model discussed above. The warm absorber ionization is
unconstrained during the {\it Swift} and XMM~1 observations. We then
fix this parameter to that obtained during the higher quality XMM~2
observation which (according to our best--fitting model, see below)
has a similar intrinsic luminosity. We force the warm absorber column
density to be the same in all observations, as in the analysis of the
high--quality {\it Suzaku} and XMM~2 observations. The joint fit is
excellent with $\chi^2 = 3140$ for 2933 dof. The most remarkable
result is that the lowest--flux XMM~1 observation appears to be
heavily absorbed by neutral gas with a column density of $\simeq
10^{23}$~cm$^{-2}$ covering about 80 per cent of the X--ray
source. Removing the absorber from the XMM~1 model, i.e. attempting to
explain the XMM~1 spectral shape with no intrinsic absorption (beside
the warm absorber), yields to a disc--reflection dominated model but
results in a poorer statistical description( $\chi^2 = 3162$ for 2936
dof). Moreover, the photon index is too flat to be realistic ($\Gamma
\sim 1.5$). We conclude that the spectral shape during the XMM~1
observation is driven by cold absorption rather than by an extremely
low intrinsic flux.

Some residuals are seen at $\sim 0.9$~keV during the lowest flux
XMM~1 observation. Adding a further soft (unresolved) X--ray emission
line at $0.904$~keV (corresponding to Ne~\textsc{ix} Ly) yields
$\Delta\chi^2 = -34$ for 1 more free parameter (although it was not
detected in our analysis, the line intensity is consistent with the
RGS spectrum of the XMM~2 observation, and it is included with the
same intensity of $1.8\times 10^{-5}$~ph~s$^{-1}$~cm$^{-2}$ in all
observations, as is the case for all other unresolved soft X--ray
emission lines, see Table~\ref{RGS}). The soft X--ray emission lines
we include in our model are most likely associated with emission from
extended gas that is photo--ionized by the AGN. The same gas may also
be associated with a soft scattered component, as ubiquitously seen in
the X--ray spectra of heavily absorbed AGN (e.g. Bianchi et
al. 2005). We test this scenario by adding to our model a soft power
law that is only absorbed by the Galactic column density. Its
normalization is the same in all spectra, as this component is not
expected to vary. The fit does not improve ($\Delta\chi^2 =-3$ for 2
more free parameters). We measure a a photon index $\Gamma \simeq 2.4$
and a scattered 0.5--2~keV luminosity of $\sim
10^{41}$~erg~s$^{-1}$. This component is overwhelmed by the X--ray
continuum, and thus negligible, in all observations except the
lowest--flux XMM~1 observation. For consistency with the typical soft
X--ray spectrum of heavily obscured AGN, we keep this component in our
best--fitting spectral model although, as mentioned, it is not
formally required by the data.

We conclude that most of the spectral variability seen in the upper
panel of Fig.~\ref{HQ4} is due to a variable cold absorber changing in
both column density and covering fraction, although photon index
variability is also detected with $\Delta\Gamma \sim 0.2-0.3$. The
most important absorption variability event is detected during the
XMM~1 observation which requires a much higher level of neutral X--ray
absorption than any other observation and, in particular, than the
{\it Swift} observation which was performed only 63 days earlier. We
measure a column density of $\simeq 10^{23}$~cm$^{-2}$ covering
$\simeq 80$ per cent of the X--ray emitting regions during the XMM~1
observation. Such a column density is more than one order of magnitude
higher than during any other observation.

\begin{figure}
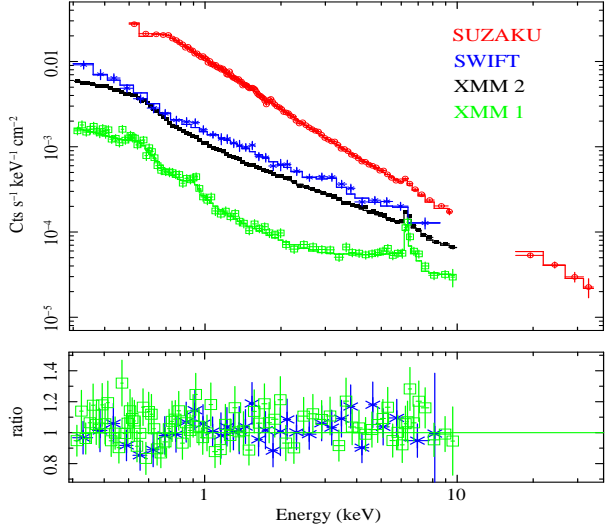

\begin{center}
\includegraphics[width=0.25\textwidth,height=0.45\textwidth,angle=-90]{h4scatt.ps}
{\vspace{0.2cm}}
\includegraphics[width=0.13\textwidth,height=0.45\textwidth,angle=-90]{h4scattratio.ps}
\caption{In the upper panel, we show the data and best--fitting models
  for the first four observations of our campaign (see
  Table~\ref{obslog}). Each spectrum is re--normalized to the
  effective area of the corresponding detector. From top to bottom, we
  show the {\it Suzaku}, {\it Swift}, XMM~2, and XMM~1 data. In the
  lower panel, we show the best--fitting data--to--model ratios for
  the {\it Swift} and XMM~1 observations. The corresponding plot for
  the {\it Suzaku} and XMM~2 observations has already been shown in
  the lower panel of Fig.~\ref{HQspec}, and it is not reproduced here
  for visual clarity.) }
\label{HQ4}
\end{center}
\end{figure}

\subsection{A hard scattered component}

As discussed extensively by Miniutti et al. (2014), the inclusion of a
hard scattered component describes very well the absorbed X--ray
states of another source which exhibits extreme X--ray absorption
variability, namely ESO~323-G77. A hard scattered component is likely
present in all cases where absorption is due to a clumpy rather than
homogeneous structure. This is because our line--of--sight (LOS) is
absorbed by a specific clump (cloud), but the X--ray continuum in
different directions intercepts different clumps which re--direct part
of it into our LOS via scattering. The presence of a hard scattered
component has also been reported in other absorbed AGN such as
NGC~3227 (Lamer, Uttley \& McHardy 2003) and NGC~7582 (Piconcelli et
al. 2007). Both sources exhibit variable X--ray absorption, most
likely associated with clumpy absorbers, in line with the scattering
interpretation outlined above. The X--ray absorption variability
ESO~362--G18 strongly suggests that this is the case here as well.

In order to check for the presence of a hard scattered component in
ESO~362--G18, we use the same phenomenological model used by Miniutti
et al. (2014) for the case of ESO~323--G77. We include an additional
absorbed power law with the same (observation--dependent) photon index
and normalization as the intrinsic nuclear continuum, we allow the
column density towards this component to be different than that
affecting the nuclear continuum (but the same in all observations),
and we multiply the model by a constant allowed to vary between 0 and
1 which represents the hard scattering fraction. The statistical
description of the data improves significantly by $\Delta\chi^2 = -28$
for 2 more free parameters for a final result of $\chi^2= 3075$ for
2928 dof. The hard scattered component is absorbed by a column density
of $\sim 2-3\times 10^{22}$~cm$^{-2}$, and its scattering fraction is
$\sim 12$ per cent. Note that this component is significantly detected
only during the most heavily absorbed XMM~1 observation (as expected,
see the discussion in Miniutti et al. 2014). The inclusion of that
component induces a higher column density and covering fraction during
the XMM~1 observation, while not affecting any other parameter. We now
measure a column density of $\sim 3-4\times 10^{23}$~cm$^{-2}$,
consistent with a full coverage of the X--ray emitting region with
covering fraction $C_{\rm f} \geq 0.94$ during the XMM~1
observation. Note that, as our phenomenological description of the
hard scattered component is likely only a rough approximation of the
true spectral shape, these values are likely affected by larger
uncertainties than the quoted statistical ones.

\begin{figure}
\begin{center}
\includegraphics[width=0.33\textwidth,height=0.45\textwidth,angle=-90]{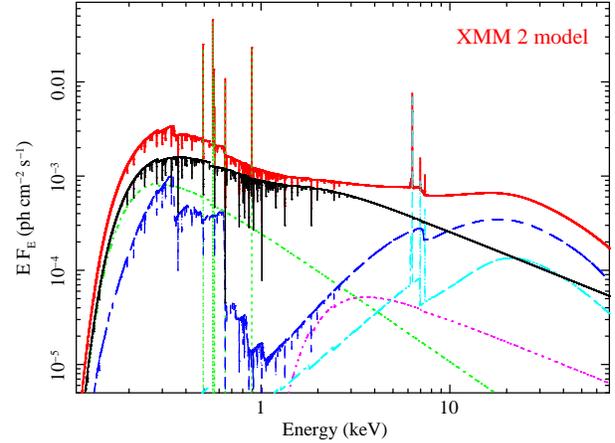}
\caption{The best--fitting model for the XMM~2 observation (upper
  solid line, red in the on--line version) together with all spectral
  components. The black solid line is the intrinsic power law,
  affected by both warm and cold (partially covering) absorption. The
  disc--reflection component is the dashed line (dark blue in the
  on--line version), while the dash--dotted line represents the
  distant reflection component (light blue in the on--line
  version). Dotted lines show the scattered components, comprising a
  soft scattered power law plus emission lines (green in the on--line
  version), and an absorbed, hard scattered component that has a
  luminosity of $\sim 12$ per cent that of the intrinsic
  continuum. Note that the latter component has a negligible
  contribution in the XMM~2 observation and that it is significantly
  detected only in the most absorbed XMM~1 observation, as discussed
  in the text.}
\label{XMM2model}
\end{center}
\end{figure}

Re--running the error contours on the relativistic parameters (see
Fig.~\ref{RelParams}) gives identical results. This is expected since
the relativistic parameters are best constrained from the
highest-quality X--ray spectra that are basically unaffected by the
inclusion of the hard scattered component. The final best--fitting
results are reported in Table~\ref{METable} and a summary and brief
discussion of the main results is presented in Section~\ref{ME}. The
data--to--model ratios for the {\it Swift} and XMM~1 observations are
shown in the lower panel of Fig.~\ref{HQ4} (see Fig.~\ref{HQspec} for
the {\it Suzaku} and XMM~2 observations). As an example of the overall
spectral model, the best--fitting model for the mildly absorbed XMM~2
observation is shown in Fig.~\ref{XMM2model} together with all
spectral components.

\begin{figure}
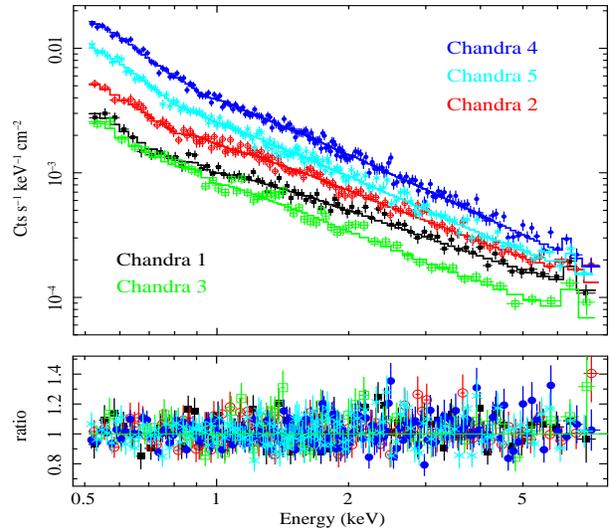

\begin{center}
\includegraphics[width=0.25\textwidth,height=0.45\textwidth,angle=-90]{CHspec.ps}
{\vspace{0.2cm}}
\includegraphics[width=0.13\textwidth,height=0.45\textwidth,angle=-90]{CHratio.ps}
\caption{In the upper panel, we show the data and best--fitting models
  for the five {\it Chandra} monitoring observations performed within
  15 days from 2010--05--18 to 2010--06--03. In the lower panel, we
  show the best--fitting data--to--model ratios resulting from the
  joint fit to the five data sets. }
\label{ch}
\end{center}
\end{figure}

\subsection{The two weeks {\it Chandra} monitoring campaign}
\label{chandra}

We consider here a joint fit to the five {\it Chandra} monitoring
observations performed between 2010--05--18 to 2010--06--03 using the
same X--ray spectral model described above. We fix the parameters
associated with the soft X--ray emission lines, the distant reflection
component, and the soft and hard scattered components to the values
derived from the analysis discussed above, as the {\it Chandra} data
have typically lower spectral quality, and because these components
are not expected to vary. The relativistic parameters affecting the
disc reflection component are also fixed to the best--fitting values
obtained from the analysis of the first four observations, as the {\it
  Chandra} data cannot constrain them any better. As before, we force
the warm absorber column density to be the same in all observations,
while the ionization parameter is free to vary independently. As for
the neutral partial covering absorber, the column density is initially
free to vary independently. However, as $N_{\rm H}$ turns out to be
consistent with being the same in all {\it Chandra} observations, we
force it to have the same value in all data sets, so that the cold
absorber is only allowed to vary in covering fraction.

We reach a good statistical description of the five {\it Chandra}
monitoring observations with a final result of $\chi^2=1248$ for 1218
dof. The spectra, best--fitting models, and final data--to--model
ratios for the joint fit to the five data sets are shown in
Fig.~\ref{ch}. The observed spectral variability can be explained by a
combination of photon index variability ($\Delta\Gamma \simeq 0.3$,
although the relatively large 90 per cent uncertainties imply that
$\Gamma = 1.9-2.0$ is an acceptable solution for all observations) and
by moderate changes in the covering fraction of a cold absorber with
column density of $\simeq 8\times 10^{21}$~cm$^{-2}$. Our
best--fitting results are reported in Table~\ref{METable} and
discussed in Section~\ref{ME}. We conclude that absorption
variability, if present at all, is only modest during the two weeks
{\it Chandra} monitoring campaign. Hence, the shortest timescale
absorption variability event we detect is that occurring within the
two months separating the {\it Swift} and the XMM~1 observations.

\section{Summary of the multi--epoch spectral analysis}
\label{ME}

\begin{table*}
\caption{Best--fitting parameters for the multi--epoch analysis. The
  constant continuum components (upper part of the Table) comprise the
  soft and hard scattered power law as well as the distant reflection
  component. $L_{0.5-2}^{\rm{(ss)}}$ is
  the luminosity of the soft scattered (ss) component, comprising
  contributions from the soft scattered power law and from all
  unresolved soft X--ray emission lines. $N_{\rm H}^{\rm{(hs)}}$ and
  $f^{\rm{(hs)}}$ are the column density towards the hard scattered
  component and its scattering fraction. The superscript
  $^{\rm{(XMM2)}}$ means that the parameter is tied to that during the
  XMM~2 observation. Luminosities are unabsorbed and given in units of
  $10^{42}$~erg~s$^{-1}$; column densities are given in units of
  $10^{22}$~cm$^{-2}$; the ionization parameter is given in units of
  erg~cm~s$^{-1}$.}
\label{METable}      
\begin{center}
\begin{tabular}{l c c c c c c c c }
\hline\hline                 
 \multicolumn{9}{c}{Constant continuum components and common relativistic parameters} \\
\hline
&  \multicolumn{2}{c}{Soft scattering} & \multicolumn{2}{c}{Hard scattering} & \multicolumn{1}{c}{Distant reflection} & \multicolumn{3}{c}{Disc--reflection relativistic parameters } \\ \\
& $\Gamma^{\rm{(ss)}}$ &$L_{0.5-2}^{\rm{(ss)}}$ & $N_{\rm H}^{\rm{(hs)}}$ & $f^{\rm{(hs)}}$ & $L^{\rm{(ref)}}_{2-10}$  &  $a$ & $q$ & $i$  \\ \\
& $2.4\pm 0.1$ &$0.12\pm 0.05$ & $2.5\pm 1.6$ & $0.12^{+0.08}_{-0.06}$ & $0.32\pm 0.09$  & $\geq 0.98$ & $4.3^{+0.8}_{-0.6}$ & $53\pm 5$  \\
\hline
\multicolumn{9}{c}{Variable components}\\
\hline
Obs.& \multicolumn{2}{c}{Cold absorber} & \multicolumn{2}{c}{Warm absorber} & \multicolumn{2}{c}{Continuum} & \multicolumn{2}{c}{Disc--reflection}\\ \\
& $N_{\rm H}^{\rm{(cold)}}$ & $C_{\rm f}^{\rm{(cold)}}$ & $N_{\rm H}^{\rm{(ion)}}$ & $\log\xi$ & $\Gamma$ & $L_{2-10}^{\rm{(nuc)}}$ & $\xi^{\rm{(ref)}}$ & $L^{\rm{(ref)}}_{2-10}$ \\ \\
\hline
Swift    & $0.5\pm 0.4$ & $0.5\pm 0.3$     & $0.15\pm 0.04$ & $2.0^{\rm{(XMM2)}}$      & $2.0\pm 0.4$   & $2.0\pm 0.3$ & $\leq 25$    & $\leq 2.0$\\
XMM~1    & $35\pm 8$    & $\geq 0.94$      & ''             & $2.0^{\rm{(XMM2)}}$      & $1.90\pm 0.09$ & $2.3\pm 0.2$ & $\leq 32$    & $0.9^{+0.5}_{-0.7}$\\
Suzaku   & $\leq 3$      & $\leq 0.1$      & ''             & $2.4\pm 0.2$ & $2.12\pm 0.03$ & $8.35\pm 0.08$ & $16\pm 12 $ & $1.9\pm 0.2$\\
XMM~2    & $1.3\pm 0.2$  & $0.42\pm 0.04$  & ''             & $2.0\pm 0.3$ & $1.80\pm 0.04$ & $1.95\pm 0.06$ & $\leq 8$    & $0.9\pm 0.1$\\
\hline
\multicolumn{3}{l}{$\chi^2/{\rm{dof}} = 3075/2928$} & & & & & &\\
\hline
Chandra~1 & $0.8\pm 0.2$ & $0.6\pm 0.1$    & $0.20\pm 0.08$ & $2.0\pm 1.1$ & $1.7\pm 0.2$   & $2.3\pm 0.2$ & $\leq 26$   & $0.7\pm 0.5$\\
Chandra~2 & ''           & $0.4\pm 0.1$    & ''             & $1.9\pm 0.6$ & $1.8\pm 0.2$   & $3.1\pm 0.1$ & $\leq 25$   & $0.9\pm 0.4$\\
Chandra~3 & ''           & $0.5\pm 0.2$    & ''             & $2.5\pm 1.3$ & $1.8\pm 0.3$   & $1.2\pm 0.4$ & $\leq 35$   & $\leq 1.6$\\
Chandra~4 & ''           & $0.48\pm 0.06$  & ''             & $2.5\pm 0.4$ & $2.1\pm 0.1$   & $4.9\pm 0.1$ & $7\pm 6$    & $1.3\pm 0.3$\\
Chandra~5 & ''           & $0.4\pm 0.1$    & ''             & $2.1\pm 0.4$ & $2.0\pm 0.1$   & $3.4\pm 0.1$ & $\leq 19$   & $1.1\pm 0.2$\\
\hline
\multicolumn{3}{l}{$\chi^2/{\rm{dof}} = 1248/1218$} & & & & & &\\
\hline\hline                 
\end{tabular}
\end{center}
\end{table*}

The relevant best--fitting parameters of our multi--epoch spectral
analysis are reported in Table~\ref{METable}. The parameters
associated with the soft emission lines, the soft/hard scattering
components, the distant reflector, and the disc--reflection
relativistic kernel are the same in all observations. They are
reported in the upper part of the Table. As mentioned, these
parameters have been fixed in the joint analysis of the {\it Chandra}
observations, so that the reported values and errors are drawn from
the joint fit to the {\it Swift}, XMM~1, {\it Suzaku}, and XMM~2
observations. 

A disc--reflection component is detected in all observations, except
the {\it Swift} and {\it Chandra}~3 ones, where only upper limits are
obtained. This is likely because of the relatively low spectral
quality of these observations (see Table~\ref{obslog}). The
relativistic parameters are best constrained from the highest--quality
spectra ({\it Suzaku} and XMM~2) and suggest that ESO~362--G18 is
powered by a very rapidly spinning Kerr black hole (with spin $a\geq
0.98$ at the 90 per cent confidence level). The observer inclination
with respect to the disc axis is also very well constrained ($i\sim
53^\circ$). The disc ionization state is rather low, and we are only
able to obtain upper limits on the ionization parameter, except during
the highest flux {\it Suzaku} and {\it Chandra}~4 observations.

The intrinsic photon index $\Gamma$ is variable and it appears to be
steeper during the highest flux observations ({\it Suzaku} and
{\it Chandra~4}), consistent with a steeper--when--brighter
behaviour. After checking that the data allow us to do so, the warm
absorber column density is forced to be the same during the first four
and during the subsequent five observations. The final results
indicate that the warm absorber column density is in fact likely
constant during the whole 2005--2010 campaign with $N_{\rm H} \simeq
1-2\times 10^{21}$~cm$^{-2}$. On the other hand, the warm absorber
ionization is marginally variable and roughly consistent with
responding to the intrinsic variability although the relatively large
errors do not allow us to investigate its variability in detail.

A neutral, partially covering absorber is detected in all observations
but the {\it Suzaku} one, which appears to be unabsorbed. The absorber
is variable, and the X--ray spectrum of ESO~362--G18 changes from
unabsorbed or only mildly absorbed states (e.g. during the {\it Swift}
and {\it Suzaku} observation) to highly absorbed ones (the XMM~1
observation), going also through states with intermediate levels of
absorption (the remaining XMM~2 and {\it Chandra} observations). The
most remarkable absorption variability event is that occurring over
the 63~days separating the mildly absorbed {\it Swift} observation
($N_{\rm H}\sim 5\times 10^{21}$~cm$^{-2}$ and $C_{\rm f} \simeq 0.5$)
and the heavily absorbed XMM~1 one ($N_{\rm H}\simeq 3-4\times
10^{23}$~cm$^{-2}$ and $C_{\rm f} \geq 0.94$).

\section{The black hole mass and further, independent constraints on the system inclination}

Garcia--Rissmann et al. (2005) measure the stellar velocity dispersion
of ESO~362--G18 from the Calcium triplet obtaining $\sigma_* = 130 \pm
4$~km~s$^{-1}$ (we consider the average value and standard deviation
of their two measurements). The black hole mass can be estimated from
the $M_{\rm{BH}}-\sigma_*$ relation (e.g. Merritt \& Ferrarese 2001;
Tremaine et al. 2002). Using the $M_{\rm{BH}}-\sigma_*$ relationship
as obtained by Xiao et al. (2011), we estimate a black hole mass of
$M_{\rm{BH}} = (0.8-1.7)\times 10^7~M_\odot$. However, in order to
include the spread and not only the statistical uncertainties on the
best--fitting $M_{\rm{BH}}-\sigma_*$ relation, we prefer to consider
the sample of reverberation--mapped AGN from Woo et al. (2010) that
have stellar velocity dispersion consistent with that of ESO~362--G18,
and to assign to our AGN the black hole mass range that can be derived
from this sub--sample. With this prescription, we obtain a larger mass
range of $M_{\rm{BH}} = (0.7-7.2)\times 10^7~M_\odot$.

Another estimate of the black hole mass can be obtained by combining
the observed optical luminosity with the H$\beta$ broad emission line FWHM
by using, e.g. the relationship from Park et al. (2012), i.e. 
\begin{equation}
M_{\rm{BH}} =
A~\left(\frac{{\rm{FWHM}}_{\rm{H\beta}}}{10^3~{\rm{km~s^{-1}}}}\right)^{1.666}
\left(\frac{\lambda
  L_{\rm{5100}}}{10^{44}~{\rm{erg~s^{-1}}}}\right)^{0.518} \, ,
\label{MBH}
\end{equation}
where $A=10^{6.985}~M_\odot$. Bennert et al. (2006) report a H$\beta$
FWHM$=5240\pm 500$~km~s$^{-1}$ from the optical spectrum obtained at
the New Technology Telescope (NTT) on 2004--09--17. They also measure
a nuclear $5100$~\AA\ luminosity of $\sim 4.1\times
10^{43}$~erg~s$^{-1}$. By using Eq.~\ref{MBH}, we then have
$M_{\rm{BH}} = (0.8-8.1)\times 10^8~M_\odot$, which appears to be
inconsistent (and much higher) than that derived from the stellar
velocity dispersion.

However, Eq.~\ref{MBH} assumes an average virial coefficient $\log
f=0.72$, as derived by Woo et al. (2010). In fact, $f$ is likely
source--dependent, and depends on the unknown BLR geometry. If we
assume a disc--like BLR geometry, the virial relationship can be
expressed in terms of line FWHM as 
\begin{equation}
M_{\rm{BH}} = R_{\rm{BLR}}
{\rm{FWHM}}^2 \left(4G\sin^2 i \right)^{-1}  \, ,
\label{MBH2}
\end{equation} 
where $i$ is the inclination between the
LOS and the angular momentum of the disc--like BLR, whose direction is
likely parallel to that of the accretion flow angular
momentum. $R_{\rm{BLR}}$ can be estimated from the Bentz et al. (2009)
BLR--luminosity relationship which, for the given
$5100$~\AA\ luminosity gives $R_{\rm{BLR}} \sim 5.2\times
10^{16}$~cm. Our disc--reflection model provides information on the
system inclination, namely $i=53^\circ \pm 5^\circ$. Inserting
$R_{\rm{BLR}}$ and $i$ into the virial relationship leads to a black
hole mass estimate of $M_{\rm{BH}} = (4.5\pm 1.5)\times 10^7~M_\odot$,
which is now totally consistent with the mass estimated from the
stellar velocity dispersion. Notice that, in order for the black hole
mass estimates from $\sigma_*$ and from the virial assumption to be
consistent with each other, one must have $\sin^2 i\geq 0.46$. In
other words, under the assumption of a disc--like BLR geometry, $i\geq
43^\circ$. This provides further, independent support to the
relatively high inclination we derive from the X--ray spectral
analysis with the disc--reflection model. Hereafter, we assume that
$M_{\rm{BH}} = (4.5\pm 1.5) \times 10^7~M_\odot$ and we refer to
$M_{\rm{best}} = 4.5\times 10^7~M_\odot$.

\section{Looking for a reverberation time lag}

An inescapable consequence of interpreting the soft X--ray excess as
partially ionized X--ray reflection off the inner accretion disc is
that relatively short time delays are expected between X--ray energy
bands dominated by the intrinsic power law, and bands dominated by
disc--reflection (i.e. the soft excess). Such lags have been indeed
discovered in the Narrow--Line Seyfert~1 galaxy 1H~0707--495 (Fabian
et al. 2009), and subsequently confirmed to be present in many other
sources (Emmanoulopoulos et al. 2011; de Marco et al. 2011; 2013; Kara
et al. 2013a, 2013b; Cackett et al. 2013; Fabian et al. 2013). The
interpretation of the soft X--ray lags as due to the delayed
reflection off the inner accretion disc has then been confirmed thanks
to the detection of very similar Fe K lags in some sources (Zoghbi et
al. 2012; Kara et al. 2013c, Zoghbi et al. 2013), although other
authors consider the detected soft lags spurious and thus physically
meaningless (e.g. Miller et al. 2010; Legg et al. 2012).

\begin{figure}
\begin{center}
\includegraphics[width=0.33\textwidth,height=0.45\textwidth,angle=-90]{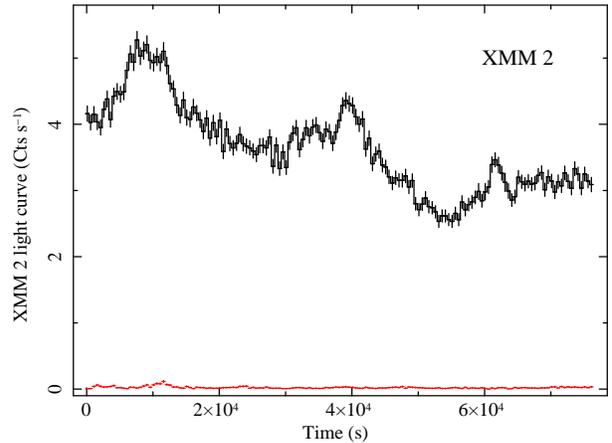}
\caption{The broadband, 0.3--10~keV EPIC--pn light curve of
  ESO~362--G18 during the XMM~2 observation is shown with a bin size
  of 500~s. The corresponding background light curve is also shown for
  comparison.}
\label{03to10lc}
\end{center}
\end{figure}

\begin{figure}
\begin{center}
\includegraphics[width=0.33\textwidth,height=0.45\textwidth,angle=-90]{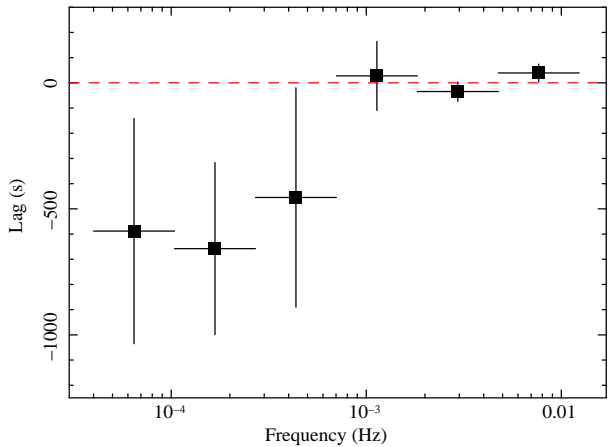}
\caption{The lag--frequency spectrum of ESO~362--G18 between the
  0.3--0.6~keV and the 0.8--3~keV bands. The lag is defined such that
  negative values mean that the softer (soft--excess--dominated) band lags the
  harder (likely continuum--dominated) one. }
\label{lag}
\end{center}
\end{figure}

ESO~362--G18 is X--ray variable during the long XMM~2 observation, and
the broadband 0.3--10~keV light curve is shown in
Fig.~\ref{03to10lc}. The X--ray variability is of sufficiently high
amplitude that lags between different energy bands can in principle be
computed. To compute the lag--frequency spectrum, we select energy
bands that, according to our best--fitting spectral model, are
dominated by disc--reflection (0.3--0.6~keV) and by the intrinsic
power law continuum (0.8--3~keV) respectively. In Fig.~\ref{lag}, we
show the lag--frequency spectrum of ESO~362--G18 between the two
selected energy bands, where negative (soft) lags mean that the soft
band lags the hard one. We measure a soft lag of $|~\tau~|= 658\pm
342$~s at $\nu \sim 1.7\times 10^{-4}$~Hz. A similar lag is consistent
to be present in the wide $0.4-7\times 10^{-4}$~Hz frequency band. For
$M_{\rm{BH}} = M_{\rm{best}}$, and assuming that soft lags are related
to the light--crossing--time between the primary and reprocessed
X--ray emission component (i.e. the power law continuum and the soft
excess), the observed soft lag translates into a distance of $(3.0\pm
1.5)~r_g$ between the continuum and reflection emission
sites. Dilution effects may slightly increase the intrinsic lag and
thus the inferred distance between continuum and reprocessing sites
(see e.g. Wilkins \& Fabian 2013). By using the best--fitting spectral
model discussed earlier, the fraction of
the reflection and power law components in the two selected bands can
be computed, and thus dilution effects can be taken into
account. Using the same arguments of Wilkins \& Fabian (2013), we
infer an intrinsic lag of $\sim 1600$~s, corresponding to a
distance of $\sim 7~r_g$ between the continuum and reflection
emitting regions.

As mentioned, the latter result depends on the physial interpretation
of the observed lag as a signature of reverberation. However, other
interpretations (based on scattering and absorption) have been
proposed (e.g. Miller et al. 2010). We cannot exclude that absorption
variability contaminates our lag measurement, although the inspection
of hardness ratios between different energy bands during the XMM2
observation, implies that spectral variability is confined within
5--10~\%, i.e. absorption variability is only modest on short
timescales.

\section{Origin of the variable absorber and constraints on the X--ray emitting region(s) size}

The main result of our X--ray absorption variability analysis is
that ESO~362--G18 makes a transition from a mildly absorbed state with $N_{\rm
  H} \simeq 5\times 10^{21}$~cm$^{-2}$ and $C_{\rm f}\simeq 0.5$ as
observed during the {\it Swift} observation, to a highly absorbed
one with $N_{\rm H} \simeq 3-4\times 10^{23}$~cm$^{-2}$ and $C_{\rm
  f}\geq 0.94$ during the XMM~1 observation performed 63 days later

The observed X--ray absorption variability is most likely due to one
absorber clump (cloud) crossing our line--of--sigth during the XMM~1
observation. A clumpy absorber may be directly identified with the
clumpy obscuring torus (e.g. Nenkova et al. 2008a; Rivers et al. 2011;
Miniutti et al. 2014; Markowitz, Krumpe \& Nikutta 2014). Another
possibility is that the variable absorber is related to BLR clouds
(e.g. Lamer, Uttley \& McHardy 2003; Elvis et al. 2004; Risaliti et
al. 2005, 2007, 2009; Sanfrutos et al. 2013; Miniutti et
al. 2014). Although the BLR and clumpy torus are likely part of the
same obscuring structure (Elitzur \& Shlosman 2006), they can be
physically distinguished by their dust content. The BLR is dust--free,
which locates the BLR within the dust--sublimation radius. The same
radius is instead generally considered as representative of the inner
edge of the clumpy torus. To gain some insight on the location and
origin of the variable absorber, we consider UV data of ESO~362--G18
with the goal of distinguishing between dust--free and dust--rich
absorption during the XMM~1 observation.

\subsection{UV variability}

If the $3-4 \times 10^{23}$~cm$^{-2}$ absorber affecting the XMM~1
observation is associated with the dusty clumpy torus, the UV 
emission from ESO~362--G18 may also be affected, provided that the
absorber covers a sufficiently high fraction of the UV--emitting
region (likely the accretion disc). On the other hand, if the absorber
is dust--free (e.g. the BLR clouds), no significant effect is expected
in the UV. Reliable optical and UV fluxes can be obtained from the
UVOT and OM telescopes on board {\it Swift} and {\it XMM--Newton}
during the mildly absorbed (hereafter {\it ``unabsorbed''} for
simplicity) {\it Swift} and XMM~2 observations as well as during the
absorbed XMM~1 one. The resulting optical/UV flux densities are given in
Table~\ref{OM}. 

The U and UVW1 fluxes are nearly consistent with being constant in all
observations. On the other hand, the UVM2 flux is nearly the same
during the unabsorbed {\it Swift} and XMM~2 observation, but $\sim 30$
per cent lower during the X--ray absorbed XMM~1 observation. The same
is true for the UVW2 flux, which is $\sim 40$ per cent lower during
the X--ray absorbed XMM~1 observation than during the unabsorbed {\it
  Swift} one (no data were collected in UVW2 during the XMM~2
observation).

It is worth pointing out that the UV intrinsic variability is an
unlikely explanation for the observed variability in the UVM2 and UVW2
filters because the intrinsic X--ray luminosity is higher during the
X--ray absorbed XMM~1 observation than during the two X--ray
unabsorbed ones. The most natural interpretation of the observed UV
variability is then that the UV--emission at the shorter wavelength
UVM2 and UVW2 filters is affected by absorption (as are the X--rays)
during the XMM~1 observation, while it is not during the two X--ray
unabsorbed ones. By comparing the UV flux during absorbed/unabsorbed
observations, and assuming that total coverage would completely block
the UV emission, we infer that the absorber covers $\sim 30$ ($\sim
40$) per cent of the UV--emitting region in the UVM2 (UVW2) filter. In
summary, the UV data are consistent with a dusty absorber covering
about 30-40 per cent of the UV emitting region during the X--ray
absorbed XMM~1 observation. This strongly suggests that the absorber
can be identified with one clump (or cloud) of the clumpy dusty torus,
rather than with a BLR cloud.

\begin{table}
\caption{Flux densities in the UVW2 UVM2, UVW1, and U filters during
  the {\it Swift} and XMM~2 (X--ray unabsorbed) observations, and the
  XMM~1 (X--ray absorbed) one. Results are given in units of
  $10^{-15}$~erg~cm$^{-2}$~s$^{-1}$~$\AA^{-1}$. }
\label{OM}      
\begin{center}
\begin{tabular}{l c c c}
\hline\hline                 
Filter & Swift$^a$ & XMM~1$^b$ & XMM~2$^b$ \\
& (unabsorbed) & (absorbed) & (unabsorbed) \\ \\
UVW2 & $13.5\pm 0.6$ & $7.9\pm 0.4$ & $-$ \\
UVM2 & $12.2\pm 0.4$ & $8.5\pm 0.4$ & $12.4\pm 0.5$ \\
UVW1 & $11.8\pm 0.7$ & $10.6\pm 0.5$ & $9.9\pm 0.5$ \\
U    & $10.0\pm 0.5$ & $10.7\pm 0.5$ & $8.8\pm 0.4$ \\
\hline\hline                        
\end{tabular}
\end{center}
\raggedright $^a$ The {\it Swift} UVOT filters are centered at $1928$~\AA\ (UVW2), $2246$~\AA\ (UVM2), $2600$~\AA\ (UVW1), and $3465$~\AA\ (U);\\
$^b$ The {\it XMM--Newton} OM filters are centered at $2025$~\AA\ (UVW2), $2250$~\AA\ (UVM2), $2825$~\AA\ (UVW1), and $3450$~\AA\ (U).
\end{table}

\subsection{The X--ray emitting region(s) size}

As the UV are only partially covered during the XMM~1 observation with
$C_{\rm{f,~UV}} \sim 0.3-0.4$, while the X--rays are fully covered
with $C_{\rm{f,~X}} \sim 1$, the X--ray emitting region is
significantly smaller in size than the UV emitting region. This is not
highly surprising, and it is in line with the mounting evidence from
e.g. microlensing results that X--rays originate in more compact,
centrally concentrated regions of the accretion flow than optical/UV
(e.g. Morgan et al. 2008; Chartas et al. 2009; Dai et al. 2010; Morgan
et al. 2012; Mosquera et al. 2013). Further constraints on the X--ray
emitting region size can be obtained from the X--ray absorption
variability itself. As the X--ray emitting region is consistent with
being fully covered during the XMM~1 observation ($C_{\rm f}\geq
0.94$), one has $D_{\rm X} \leq D_{\rm c}$, where $D_{\rm X}$ and
$D_{\rm c}$ are the assumed linear sizes of the X--ray emitting region
and of the obscuring cloud respectively. On the other hand, $ D_{\rm
  X} = \Delta T v_{\rm c}$, where $v_{\rm c}$ is the absorbing cloud
velocity and $\Delta T \leq 63$~days is the time it takes to make the transition
from an unobscured to a fully covered spectral state\footnote{Although
  the {\it Swift} observation is mildly absorbed, the large difference
  in column density with the XMM~1 observation excludes that the same
  structure was obscuring the {\it Swift} and XMM~1
  observtions. Hence, for the purpose of our discussion, the {\it
    Swift} observation is considered to be unabsorbed.}.

As discussed above, the UV data strongly suggest that the absorbing
structure can be identified with one cloud of the clumpy, dusty
torus. Hence, $v_{\rm c}$ is lower than the Keplerian velocity at the
dust sublimation radius $R_{\rm{dust}} \sim 0.4 L_{\rm{Bol,
    45}}^{0.5}$~pc (Nenkova et al. 2008a), where $L_{\rm{Bol, 45}}$ is
the bolometric luminosity in units of $10^{45}$~erg~s$^{-1}$. The
(averaged) bolometric luminosity of ESO~362--G18 can be estimated from
the 2--10~keV luminosity assuming a X--ray bolometric correction
$k_{2-10}$. We use here the total (rather than the power law) averaged
and unabsorbed 2--10~keV luminosity $L_{2-10}^{\rm{(tot)}} \sim
5.1\times 10^{42}$~erg~s$^{-1}$, as material at $R_{\rm{dust}}$ sees
all contributions. As for the X--ray bolometric correction, we assume
$k_{2-10} = 25$ (Vasudevan et al. 2009), which gives $L_{\rm{Bol}} =
1.3 \times 10^{44}$~erg~s$^{-1}$ (i.e. an Eddington ratio of
$L_{\rm{Bol}}/L_{\rm{Edd}} \simeq 0.02$ for a black hole mass of
$M_{\rm{BH}} = M_{\rm{best}} = 4.5\times 10^7~M_\odot$). From the
estimated $L_{\rm{Bol}}$ we have that $R_{\rm{dust}} \sim
0.14$~pc. Assuming Keplerian motion of the clumpy torus clouds, one
has that $v_{\rm c} \leq 1180$~km~s$^{-1}$ for $M_{\rm{BH}} =
M_{\rm{best}}$.

\begin{figure}
\begin{center}
\includegraphics[width=0.459\textwidth,height=0.612\textwidth,angle=-180]{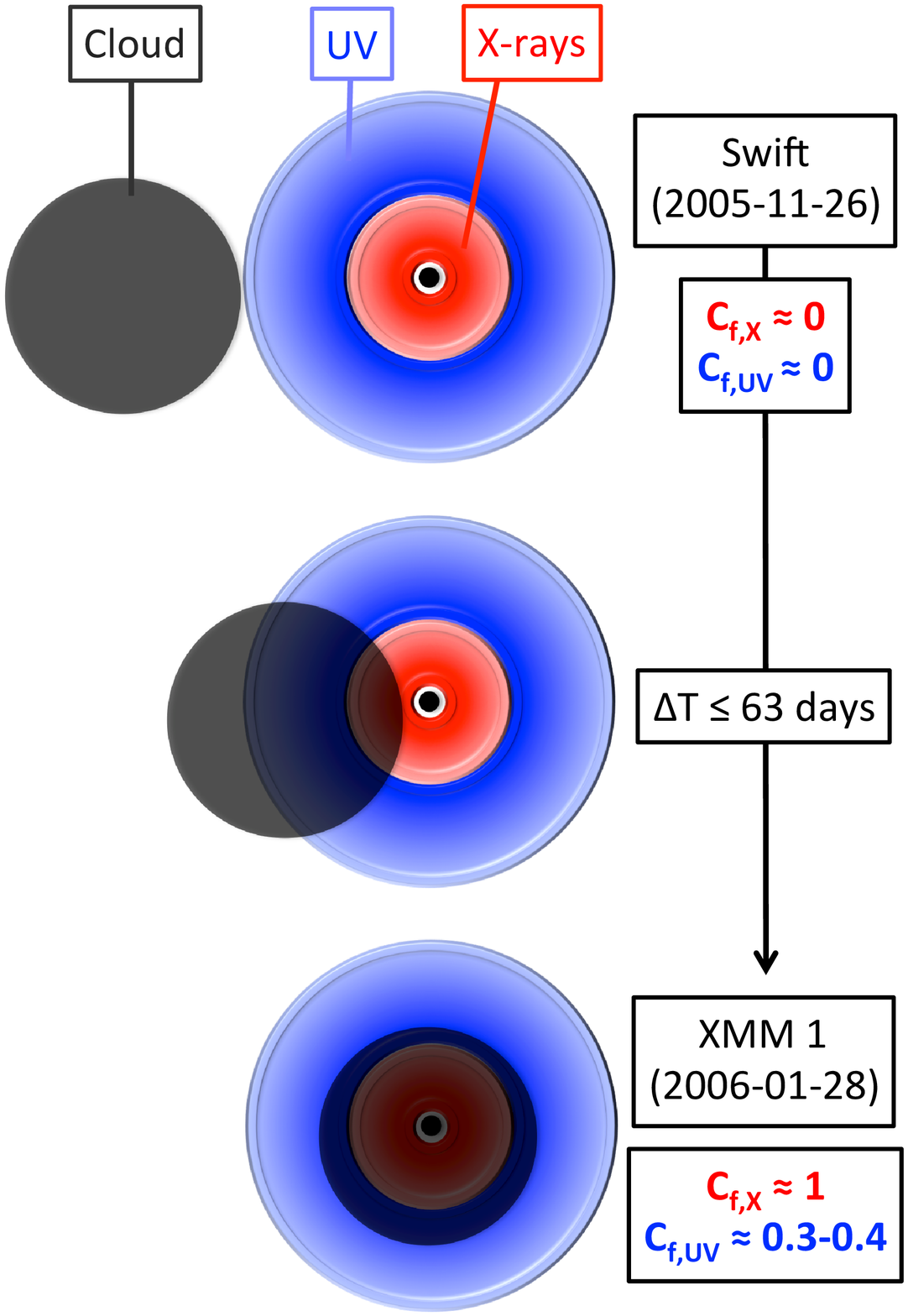}
\caption{A simple sketch of the possible system geometry as a function
  of time during the transition between the unabsorbed {\it Swift} and
  the heavily absorbed XMM~1 observation. We show the larger
  UV--emitting region (blue in the on--line version), the smaller
  X--ray one (red in the on--line version), and the absorbing cloud
  (dark gray). The figure is not to scale for visual clarity. The
  time--evolution proceeds from top to bottom. The upper panel is
  representative of the unabsorbed {\it Swift} observation. The lower
  panel represents the likely geometry during the absorbed XMM~1
  observation performed 63~days later, when the UV--emitting region is
  partially covered ($C_{\rm{f,~UV}} \sim 0.3-0.4$) by the dusty absorbing
  cloud, while the X--ray--emitting region is fully covered
  ($C_{\rm{f,~X}} \sim 1$). An intermediate situation where both the
  UV and X--ray--emitting regions are partially covered is shown in
  the middle panel.}
\label{geo}
\end{center}
\end{figure}

Substituting the upper limit on $v_{\rm c}$ into $ D_{\rm X} = \Delta
T v_{\rm c}$ gives $D_{\rm X} \leq 6.4\times 10^{14}$~cm. This
corresponds to $D_{\rm X} \leq 96~r_g~M_{\rm{best}}/M_{\rm{BH}}$, so
that X--rays come from radii within $D_{\rm X}/2 =
48~r_g~M_{\rm{best}}/M_{\rm{BH}}$ from the central, accreting black
hole under the natural assumption of axial symmetry. This is
consistent not only with microlensing results (e.g. Mosquera et
al. 2013 and references therein), but also with measurements of the
X--ray emitting size coming from other, better monitored X--ray
occultation events which imply $D_{\rm X} \leq 10-20~r_g$
(e.g. Risaliti et al. 2007, 2009; Sanfrutos et al.2013). A simple
sketch of the envisaged geometry and time--evolution is shown in
Fig.~\ref{geo}.

The upper limit on $D_{\rm X}$ also represents a lower limit on
$D_{\rm c}$. By combining this with the maximum observed column
density, we infer that the cloud number density is $n_{\rm c} \leq
6.7\times 10^8$~cm$^{-3}$. Such a density is lower than that required
for BLR clouds ($\geq 10^9$~cm$^{-3}$, e.g. Davidson \& Netzer 1979),
supporting our identification of the variable absorber with one clumpy
torus cloud.

\section{Discussion}
\label{discussion}

Multi--epoch X--ray observations ESO~362--G18 reveal that the AGN is
generally mildly absorbed by both partially ionized and
neutral/low--ionization gas. While the warm absorber properties do not
change significantly during our monitoring, the neutral absorber
exhibits remarkable spectral variability with one observation (XMM~1)
being highly significantly more absorbed than all others. 

While unabsorbed, the X--ray spectrum is characterized by a strong
soft excess, typical of the X--ray spectra of type 1 AGN. A broad
feature is also seen at Fe K energies, and the {\it Suzaku} data from
the HXD also exhibit a hard X--ray excess around 20--30~keV. All
features are highly reminiscent of a reflection component off the
inner accretion disc, whose spectral shape is distorted by
relativistic effects. Using data from the two highest--quality,
relatively unabsorbed observations {\it Suzaku} and XMM~2, we show
that the X--ray spectrum is indeed best--described by a power law
continuum affected by warm absorption, a reflection component from
distant matter (comprising an unresolved Fe emission line) and, most
importantly, a relativistically distorted disc--reflection spectrum
originating in the inner accretion flow. Although its significance is
rather low (about 2$\sigma$, see Fig.~\ref{lag}), an X--ray
reverberation lag between the continuum-- and
disc--reflection--dominated bands supports our spectral model and
suggests that the continuum and reprocessing sites are separated by
$\sim 7~r_g$. From our disc--reflection model, we infer that
the black hole powering the AGN in ESO~362--G18 is a rapidly spinning
Kerr black hole, and we measure a black hole spin $a\geq 0.92$ at the
99.99 per cent (statistical) confidence level. The disc--reflection
emissivity profile is steep, with $q=4.3^{+0.8}_{-0.6}$, and the
inclination $i$ between the disc axis and our LOS is $53^\circ \pm
5^\circ$. Such a relatively high inclination implies that our LOS may
intercept the classical obscuring torus of AGN unification schemes
whose half--opening angle is typically assumed to be of the order of
$45^\circ$. 

Indeed, we observe X--ray absorption variability during our campaign,
pointing to a clumpy nature of the cold absorber. All observations but
one are consistent with being unabsorbed, or only mildly absorbed. The
absorbed observation XMM~1 is characterized by a column density of
$\sim 3-4\times 10^{23}$~cm$^{-2}$ and a covering fraction of $C_{\rm
  f} \geq 0.94$. On the other hand, the previous {\it Swift}
observation performed only 63~days earlier is consistent with a
typical mildly absorbed state, with a neutral column density about two
orders of magnitude lower. UV variability between the absorbed XMM~1
and the unabsorbed {\it Swift} and XMM~2 observations strongly
suggests that absorption is due to a dusty cloud of the clumpy torus
transiting the LOS during the (only) heavily obscured observation of
our monitoring campaign.

It is worth pointing out that, in our best--fitting spectral model,
the absorber is applied to both the power law continuum and the
disc--reflection component with the same covering fraction in each
observation. We now perform a simple check of this assumption. We
select the two most heavily absorbed observations (namely the XMM~1
and XMM~2 ones). Our best--fitting model, obtained by assuming the
same covering fraction towards the two spectral components in each
observation, results in $\chi^2= 1634$ for 1579 dof. We then allow the
covering fraction towards the power law and the disc--reflection
component to be different. We reach a similar statistical quality
($\chi^2=1631$ for 1577 dof), and the covering fraction towards the
two components is consistent with being the same during each
observation. This means that any constraint that can be obtained on
the size of the X--ray emitting region is valid for both spectral
components.

The relatively short X--ray absorption variability timescale of $\sim
2$~months between the {\it Swift} and XMM~1 observations enables us to
constrain the X--ray emitting--region size to $D_{\rm X} \leq 96~r_g$,
i.e. assuming the natural axial symmetry, $D_{\rm X}/2 = R_{\rm X}
\leq 48~r_g$. As absorption affects both the power law continuum
(X--ray corona) and soft excess (disc--reflection, according to our
spectral decomposition), $R_{\rm X}$ has to be associated with the outer
boundary of the largest of the two emitting sites. The resulting
compact nature of the soft excess emitting--region is fully consistent
with our interpretation in terms of disc--reflection as, according to
our model (and the steep emissivity profile), the bulk of the X--ray
reflection flux originates from radii within $\sim 10~r_g$. If the
observed time lag between continuum and soft excess dominated energies
can be interpreted roughly as the light--crossing--time between the
continuum and soft excess emitting sites (a delay that would then
correspond to a distance of $\sim 7~r_g$), our upper limit on the
X-ray source size(s) is likely to significantly overestimate the real
geometrical sizes. 

As the X--ray source is consistent with being fully
covered during the XMM~1 observation, the cloud linear size must be
larger (or at most equal) to the X--ray source size. By combining this
result with the observed column density we infer a cloud density of
$n_{\rm c} \leq 6.7\times 10^8$~cm$^{-3}$. Such a density is lower
than that required for BLR clouds, supporting our
identification of the variable absorber with one cloud of the dusty,
clumpy torus.

\section{Conclusions}

In summary, ESO~362--G18 is a Seyfert galaxy characterized by an
averaged bolometric luminosity of $1.3\times 10^{44}$~erg~s$^{-1}$
which, considering our best--estimate for the black hole mass of
$4.5\times 10^7~M_\odot$, transaltes into an Eddington ratio of
$0.02$. The X--ray spectrum comprises a reflection component off the
inner accretion disc around an almost maximally spinning Kerr black
hole, and we infer a disc inclination of $\sim 53^\circ$. Such high an
inclination is expected to intercept, at least at times, the
atmosphere of the obscuring torus which is generally thought to have
an half--opening angle of $\sim 45^\circ$. Indeed, we detect variable
X--ray absorption in our data, with the most remarkable event
occurring over $\leq 2$~months. The X--ray and UV data enable us to
identify the variable absorber with the clumpy, dusty torus. The
observed occultation event suggests that both the X--ray continuum and
soft excess originate in a compact region within $\sim 50~r_g$ from
the central black hole. This supports our view that the soft excess is
due to relativistic disc--reflection off the partially ionized disc
surface. The detection of a relatively short time lag between the
continuum and the soft excess also supports this interpretation and
the inferred continuum--to--reprocessor distance ($\sim 7~r_g$) may
indicate that the actual size of the X-ray emitting region(s) is
significanlty overestimated by our upper limit.

\section*{Acknowledgments}

This work is based on observations obtained with {\it XMM--Newton}, an
ESA science mission with instruments and contributions directly funded
by ESA Member States and NASA. We made use of data obtained from the
Chandra Data Archive, and software provided by the Chandra X-ray
Center (CXC). We acknowledge the use of public data from the {\it
  Swift} data archive. This work also made use of data from the {\it
  Suzaku} observatory, a collaborative mission between the space
agencies of Japan (JAXA) and the USA (NASA). This work has made use of
data and/or software provided by the High Energy Astrophysics Science
Archive Research Center (HEASARC), which is a service of the
Astrophysics Science Division at NASA/GSFC and the High Energy
Astrophysics Division of the Smithsonian Astrophysical
Observatory. This research was funded by the European Union Seventh
Framework Program (FP7/2007--2013) under grant 312789. BAG thanks the
Spanish MINECO for support through the FPI program associated with
grant AYA2010-21490-C02-02. MS thanks the CSIC JAE-Predoc program for
support. NLS acknowledges support by ESA through the ESAC trainee
program.

\end{document}